\documentclass[twoside]{article}
\usepackage{graphicx}
\usepackage{fancyhdr}
\usepackage{multicol}
\usepackage{styl-ap}

\begin{document}
\title{Search for gravitational waves from supernovae and long GRBs}
\author{Maurice H.P.M. van Putten\work{1}}
\workplace{Korea Institute for Advanced Study, Hoegiro, Dongdaemun-gu, Seoul 130-722, Korea, {\rm and} 
Department of Astronomy, Sejong University, 98 Gunja-Dong Gwangin-gu, Seoul 143-747, Korea}
\mainauthor{mvputten@kias.re.kr}
\maketitle

\begin{abstract}
We report on evidence for black hole spindown in the light curves of the BATSE catalogue of 1491 long GRBs by application of matched filtering.
This observation points to a strong interaction of the black hole with surrounding high density matter at the ISCO,
inducing non-axisymmetric instabilities sustained by cooling in gravitational wave emission. Opportunities for
LIGO-Virgo and the recently funded KAGRA experiments are highlighted, for long GRBs with and without supernovae
and for hyper-energetic core-collapse supernovae within a distance of about 35 Mpc in the Local Universe.
\end{abstract}


\begin{multicols}{2}

\section{Introduction}

Gamma-ray bursts (GRBs) and core-collapse supernovae (CC-SNe) are by far the most energetic and enigmatic
transient events associated with neutron stars and mass black holes. A key objective is to identify the nature of their
explosion mechanism. 

Extremely powerful events are unlikely powered by their presumably prodigious output in MeV neutrinos. They may, instead, be powered by rotation (e.g. \cite{bis70}).  GRB 030329/SN 2003dh and GRB 031203/SN 2003lw are hyper-energetic events, requiring an energy reservoir that exceeds the maximal rotational energy of a proto-neutron star (PNS) by an order of magnitude
\cite{van11}. These anomalous supernovae - of ``Type I X" - practically rule out PNS as universal inner 
engine to all core-collapse supernovae. 

In light of these hyper-energetic events and the diversity in GRBs in durations and associations (with and without supernovae,
e.g., listed in \cite{van11}), we here turn to inner engines hosting a rotating black hole surrounded by high-density matter. These systems appear naturally in mergers of neutron stars with another neutron star or companion black hole, as well as in 
core-collapse of relatively mass stars. 

\subsection{Frame dragging induced outflows} 

According to the Kerr metric, the angular momentum of rotating black holes 
induces frame dragging. Frame dragging has recently been measured by the LAGEOS II \cite{ciu04} and Gravity Probe B \cite{eve11} 
satellite experiments around the Earth, manifest in an angular velocity $\omega\sim J/r^3$ at an orbital radius
$r$ and given the Earth's angular momentum $J$. By this scaling property, these experiments equivalently measured $\omega$ 
around a maximally spinning Kerr black hole at a distance of about 5.3 million Schwarzschild radii.  

As part of the gravitational field, the frame dragging angular velocity $\omega$ is a clean and universal agent inducing non-thermal radiation processes in spacetime around rotating black holes. Relativistic frame dragging is encountered as $\omega$ approaches
the angular velocity $\Omega_H$ of the black hole itself. 

Frame dragging around rapidly rotating black holes enables transfer of a {\em major} fraction of its
spin energy to surrounding matter via an inner torus magnetosphere simultaneously with transfer 
of a {\em minor} fraction to infinity in the form of ultra-relativistic outflows along the spin axis. The first can
be seen by an equivalence in poloidal cross-section of the inner and outer faces of a torus in suspended accretion to the 
magnetospheres of neutron stars \cite{van99,van03}. The second is described by a potential energy ${\cal E}=\omega J_p$ for test particles with angular momentum $J_p$. Here, ${\cal E}$ assumes energies on the order of Ultra High Energy Cosmic Rays (UHECRs) for particles in superstrong magnetic fields, typical for the catastrophic events under consideration \cite{van09}. 
Frame dragging is hereby a causal mechanism for the onset of a two-component outflow, 
comprising a baryon-rich wind from an inner disk or torus collimating the latter in the form of a baryon-poor jet. 

A two-component outflow with different baryon loading is advantageous for GRB-supernovae, whose GRB emissions
are generally attributed to dissipation of ultra-relativistic baryon-poor outflows and whose supernovae can be 
attributed to irradiation from within \cite{van03}, such as by impulsive momentum transfer of internal magnetic winds onto
the remnant stellar envelope. The efficiency of the latter favors relatively baryon-rich winds \cite{van11}, which
generally will produce aspherical explosions. These two-component outflows circumvent the limitations of 
neutron stars with approximately uniform baryon-loading throughout their wind, rendering these less amenable
to making hyper-energetic supernovae with a successful GRB. 

\subsection{If not a pulsar, how to identify rotation?}

In our model, the $T_{90}$ durations of long GRBs is attributed to the lifetime of rapid spin of the black hole. It sets 
the duration for energy and angular momentum transfer from the black hole mostly onto the surrounding matter. 
When the black hole has slowed down sufficiently, this interaction ceases, and the inner accretion disk or torus 
will fall in, heralding the end of the GRB. That is, a long GRB ends with hyper-accretion onto a slowly spinning black hole, 
whose angular velocity is approximately equal to that of the ISCO as prescribed by the Kerr metric. 

Furthermore, our model considers the aforementioned two-component outflow. Any time-variability in the collimated torus
wind - in strength or orientation - inevitably modulates the inner jet producing the GRB with possibly quasi-periodic variations 
in the light curve. The frequency scale of these variations are those associated with the ISCO around the black hole, i.e., on the
order of 1 kHz, possibly higher if multipole mass moments are involved and possibly lower by precession.
A search for such high frequency QPOs requires the highest sampling frequencies available
for GRB light curves,  well beyond the 64 ms sampling interval in the BATSE catalogue. 

Following the above, we consider the problem of tracking the evolution of black hole spin, as determined by its interaction
with a surrounding torus. Attributed to frame dragging as indicated above, the count rate of gamma-ray photons observed 
should show a near-exponential decay as the angular velocity approaches the fixed point $\Omega_H=\Omega_{ISCO}$.

The evolution of an initially rapidly rotating black hole that ensues is mostly in a decrease in its angular momentum and less 
so in a decrease of its total mass \cite{van08,van11a}. 
Since $\omega$ and hence ${\cal E}$ decrease along with $\Omega_H$, the luminosity of 
the baryon-poor outflows will decrease in the process of black hole spindown in approaching a fixed 
point $\Omega_H=\Omega_{D}$. Here, $\Omega_D$ denotes the angular velocity of the inner disk, which is
expected to closely track the angular velocity $\Omega_{ISCO}$ of the Inner Most Stable Circular Orbit (ISCO), as described 
by the Kerr metric. Thus, an asymptotically exponential decay in the light curve of any high energy emission produced by 
dissipation of the aforementioned baryon poor outflows is a characteristic imprint of black hole spindown. 

The above suggests extracting a normalized light curve (nLC) of long GRBs, to
identify the anticipated late time, exponential decay. Here, we report on this study using a recent implemented matched 
filtering \cite{van09}, to identify a model light curve in the complete BATSE catalogue of 1491 long GRBs. Matched filtering
allows for an accurate extraction of a normalized light curve, in which fluctuations on all intermediate time scales are filtered out, 
that can be used to validate model templates. We apply this to various model templates representing spindown, of black holes and
(proto-)neutron stars.

\subsection{EM priors to GW searches}

Properties of the inner engine of hyper-energetic supernovae and long GRBs in the electromagnetic spectrum
of radiation can provide valuable priors to searches for their gravitational wave emissions. By the very nature of these
catastrophic events, the high density matter at nuclear densities orbiting at about the Schwarzschild radius of the central engine,
is expected to develop a non-axisymmetric mass distribution in response to exceptionally large energy fluxes.
If so, significant emission in gravitational waves is inevitable for a sustained period of time, provided by a balance between
heating, apparent in MeV neutrino emission, and cooling, by aforementioned magnetic disk winds and gravitational radiation.
This outlook offers some novel opportunities for the advanced detectors LIGO-Virgo and KAGRA currently under construction, 
that are expected to be operational within this decade.

\begin{myfigure}
\centerline{\resizebox{70mm}{!}{\includegraphics{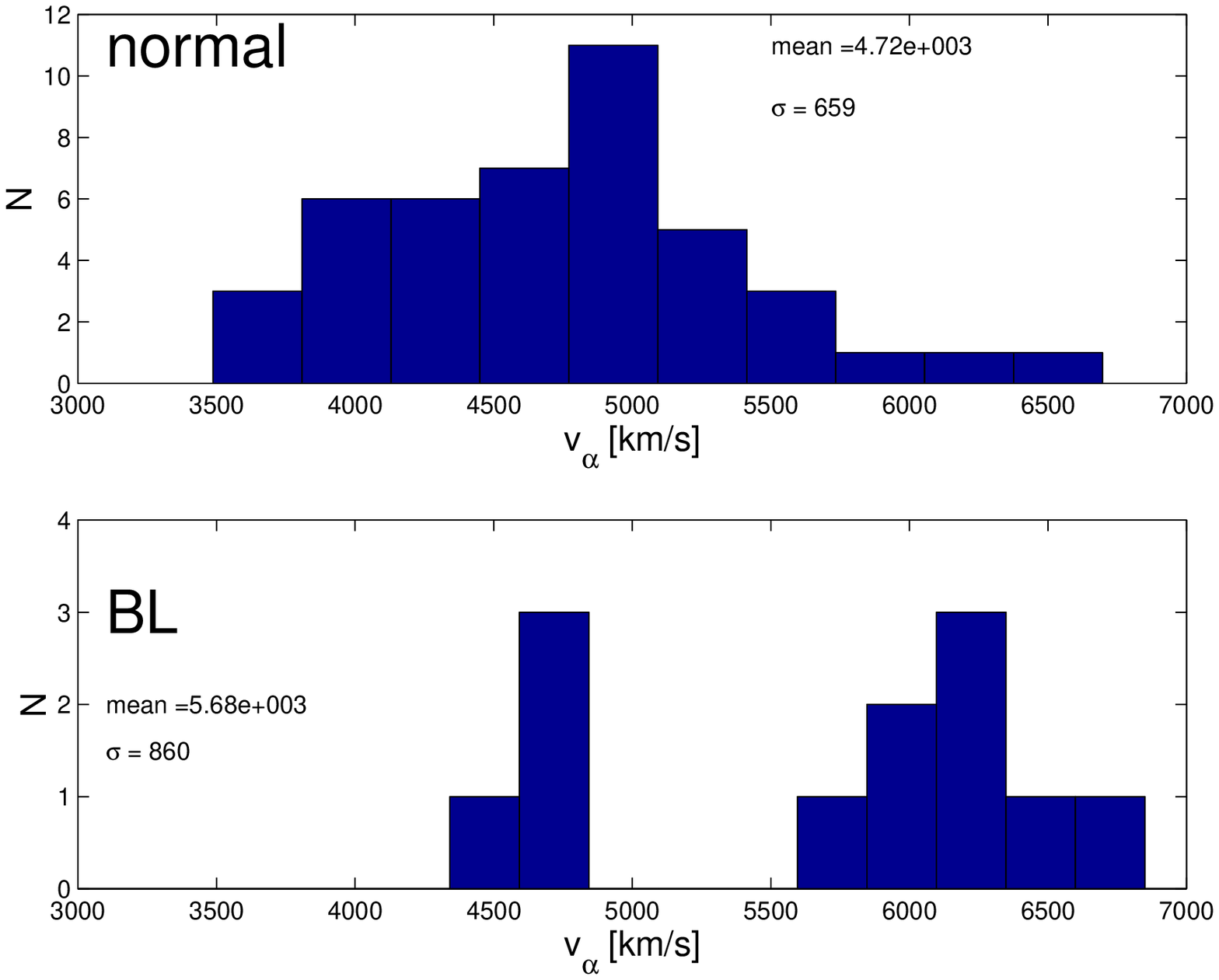}}}
\caption{Shown are the histograms of the ejection velocities of the sample of \cite{mau10} of CC-SNe with narrow ($top$) and broad ($bottom$) line emission lines. The mean of the latter is larger than that of the former by $3.9\sigma$.}
\label{author-fig1}
\end{myfigure}

Of particular interest are priors to identifying black holes or PNS, in view of their dramatically different
outlook in gravitational wave emission. The latter offers a broad range of possible radiation channels which, however, 
remain hitherto somewhat uncertain. Examples are  acoustic modes, convection, differential rotation, and magnetic fields 
\cite{ree74,owe98,cut02a,cut02b,arr03,how04,dal09,and11,reg11}.
If an electromagnetic prior is found to rule out a PNS, we can direct our 
attention to black holes as a probable alternative. An attractive possibility, for example, is a prolonged quasi-period emission by quadrupole mass inhomogeneities or non-axisymmetric instabilities in the inner disk or torus at twice the orbital frequency, particularly so if powered by the spin energy of the black hole.

\begin{mytable}
\caption{Selected supernovae with kinetic energies $E_{SN}$ in units of $10^{51}$ erg. Required energy
reservoir $E_{rot}$ is expressed in $\hat{E}=E_{rot}/E_c$, $E_c=3\times 10^{52}$ erg (adapted from \cite{van11}).}
\label{author-tab1}
\bigskip
\centerline{\begin{tabular}{|lllll|l|}
\hline
GRB			& SN		& $E_{SN}$ & $\eta$ & $\hat{E}$ & Prior\\	
\hline
  			& 2005ap 	& $>$10	& 1		& $>0.3$ 	& indet\\
  			& 2007bi  	& $>$10	& 1		& $>0.3$ 	& indet\\
  980425  	& 1998bw	& 50		& 1		& 1.7 	& BH\\
  031203	& 2003lw	& 60		& 0.25	& 10		& BH\\
  060218   & 2006aj	&  2		& 0.25	& 0.25	& indet\\
  100316D & 2010bh     & 10 		& 0.25	& 1.3		& BH\\
  030329  	& 2003dh     & 40		& 0.25	& 5.3		& BH\\
\hline
\end{tabular}}
\end{mytable}

\section{Diversity out of universality}

In contrast to Type Ia supernovae, CC-SNe are not of one kind with both narrow and broad emission lines, shown
in the histograms of ejection velocities (Fig. 1) of a sample of CC-SNe compiled by \cite{mau10}. Those of
broad emission line events are higher on average that those of narrow emission line emission events
with a statistical significance in excess of 4 $\sigma$. The explosions of the former appear to be 
relatively more energetic.

Rapidly rotating black holes $(\Omega_H>\Omega_{ISCO}$, $a/M>0.36$) can form in CC-SNe and mergers of 
two neutron stars. If the progenitor of the former is a member of a short period binary, a rapidly rotating black hole 
is formed with a rotational energy between about one- and two-thirds of the extremal value for Kerr black holes. 
In particular, dimensionless specific angular momenta $a/M=0.7679$ and $a/M=0.9541$ are found with $E_{rot}/E_{rot}^{max}=0.3554$ and, respectively, $E_{rot}/E_{rot}^{max}=0.6624$. Neutron star-neutron star mergers form rapidly rotating 
black holes of relatively small mass, $M$, essentially the sum of the mass $M_{NS}$ of individual neutron stars 
except for a minor loss of mass that forms an accretion disk, for which we estimate
\begin{eqnarray}
\frac{a}{M} =  \frac{2}{5} \sqrt{\frac{R}{R_g}} =0.5963,~0.7303,~0.8433
\end{eqnarray}
for $M_{NS}=3,~2,~1.5M_\odot$.
These values are in remarkable agreement with numerical simulations, 
showing $a/M=0.74-0.84$ for $M_{NS}=1.5M_\odot$ \cite{bai08}. In the merger of
a neutron star with black hole companion, an accretion disk will form from tidal break-up around the latter if
the black hole is not too massive (the limit for which is relaxed if it spins rapidly). In this event, the spin of 
the black hole is unchanged from its spin prior to the merger. Consequently, the rotation of the black hole
may be diverse, given by the diversity spin in neutron star-black hole binaries. 

In considering GRBs from rotating black holes, we associate long/short GRBs with rapidly/slowly spinning black holes. 
The above leads to the perhaps counter-intuitive conclusion that {\em mergers of neutron star-neutron star binaries 
produce long GRBs, more likely so than short GRBs} \cite{van08}. They may further be produced by mergers of 
neutron stars with black hole companions (e.g. \cite{pac91,klu98}), especially those with rapid spin \cite{van99}. 

Consequently, long GRBs are expected to occur both with and without supernovae involving
rapidly rotating black holes. 
Notable examples are given in Table 1 of \cite{van11}, e.g., GRB 060614 with $T_{90}=102$ s.

\section{A light curve of long GRBs}

To leading order, the evolution of rotating black hole interacting with surrounding matter via a torus magnetosphere
is described by conservation of energy and angular momentum associated with the black hole luminosity $L_H$,
its mass $M$ and angular momentum $J_H$, satisfying \cite{van99}
\begin{eqnarray}
L_H= -{\dot M},~~\dot{J_H}=-T,
\label{EQN_SD}
\end{eqnarray}
where $L_H=\Omega_HT$ associated with the torque $T=\kappa(\Omega_H-\Omega_D)$. Here, $\kappa$ 
incorporates the physical and geometrical properties of the inner torus magnetosphere, and $\Omega_D$ is 
taken to be tightly correlated to $\Omega_{ISCO}$ mentioned earlier. Upon numerical integration 
of (\ref{EQN_SD}), a model light curve can be calculated from the resulting baryon-poor outflow as a function of 
$\Omega_H=\Omega_H(t)$ and $\theta_H=\theta_H(t)$ in its dependence on the radius $R_D=R_{ISCO}$ 
of the inner disk. For our intended leading order analysis, we assume that the photon count rate is 
proportional to the luminosity in the baryon poor outflow. Fig. 2 shows the resulting template, 
scaled to a duration of about one minute. 

A finite polar cap supports an open magnetic flux tube to infinity, while the remainder of the black hole
event horizon is connected to the inner disk or torus via an inner torus magnetosphere. The time-averaged 
luminosity of the jet along the open magnetic flux tube that results from the action of frame dragging, 
satisfies the angular scaling \cite{van03}
\begin{eqnarray}
\left< L_j(t)\right> \sim 10^{51} \left(\frac{M_1}{T_{90}/30\mbox{s}}\right)\left<\left(\frac{\theta_H(t)}{0.5}\right)^4\right>\mbox{erg~s}^{-1}
\label{EQN_Lj}
\end{eqnarray}
with further dependence on the angular velocity of the black hole according to 
$L_j\propto \Omega_H^2 z^2 E_{k,T}$, $E_{k,D}\propto (\Omega_DR_D)^2 e(z)$, $e(z)=\sqrt{1-\frac{2}{3z}},$
$R_T \equiv zM$. Here, we consider the geometrical ansatz $\theta_H^4\propto z^n$ with $n=2$. (Here, $n=2$
corresponds to the expansion of the surface area $A_H=A_1z+A_2z^2+\cdots$ of the polar cap as a function of
the radius of the inner disk.) Fig. 2 shows the model light curve produced by (\ref{EQN_Lj}) following (\ref{EQN_SD}).

\begin{myfigure}
\centerline{\resizebox{70mm}{!}{\includegraphics{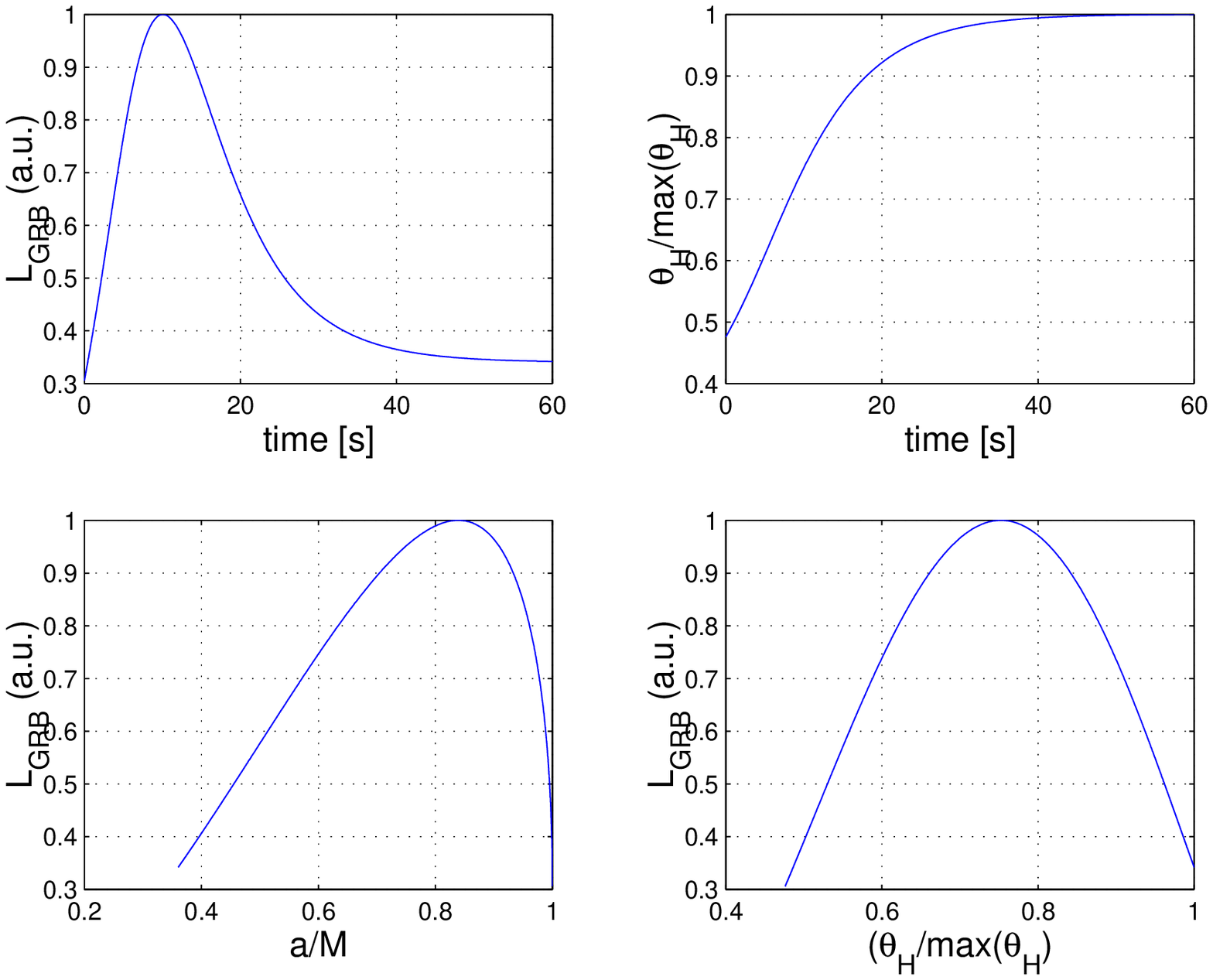}}}
\caption{The template light curve in gamma-rays for an initially extremal black hole, shown as a function of time  ($top$ $left$), dimensionless angular momentum ($bottom$ $left$), and the associated horizon half-opening angle $\theta_H$ ($right$). During initial spindown, the rise is due to a marked increase of $\theta_H$ with the expansion of the ISCO, leading to a maximal luminosity at about 16\% of the duration of the burst. There is an accompanying decrease in the total mass of the black hole \cite{van08,van11}.}
\label{author-fig2}
\end{myfigure}

Thus, the light curve (\ref{author-fig2}) is based on two positive correlations, in the GRB luminosity with $\omega$ and
$\theta_H$ with the radius of the ISCO. For the latter we use the Ansatz n=2, a choice among a few integers by insisting
on analyticity. We now seek validation of this Model A in the BATSE light curves of long GRBs by application of matched filtering.

\section{Normalized GRB light curve}

To extract an intrinsic light curve of long GRBs, we set out to filter out all fluctuations, randomly or quasi-periodic, in the 
BATSE data (Fig. 3) that are not representative for the evolution of the black hole. These include, without being exhaustive, precession of the disk or black hole and instabilities, in the accretion disk, the inner torus magnetosphere and in the interface between baryon-poor jet and the surrounding baryon-rich disk winds, as well as fluctuations in the dissipative fronts due to turbulence. To this end, a normalized light curve (nLC) is extracted by an ensemble average of individually normalized light curves defined by a best-fit to a template upon scaling and translation of time and count rate. 

\begin{myfigure}
\centerline{\resizebox{80mm}{!}{\includegraphics{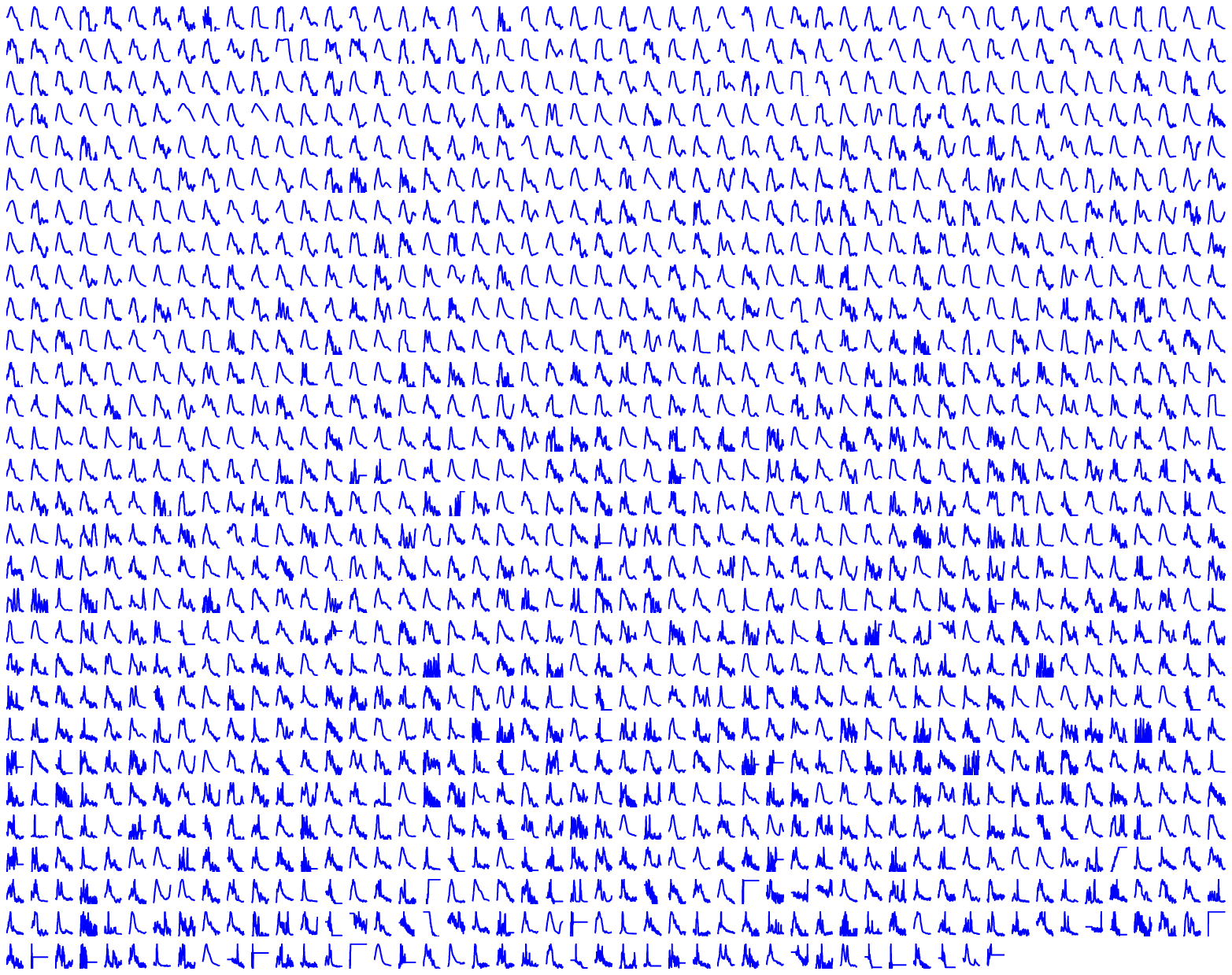}}}
\caption{Compilation of the complete BATSE catalogue of 1491 light curves of long GRBs, sorted by 2 s $<T_{90}<$ 1307 s. Each light
curve shown represents the sum of the photon count rate in all four BATSE energy channels, is smoothed with a time scale of 2.56 s and
is plotted as a function of time normalized to $T_{90}$. (Reprinted from \cite{van12}.)}
\label{author-fig3}
\end{myfigure}

Fig. 4 shows the results along with the residuals, showing convergence for very long bursts ($T_{90}>20$ s) to Template A (Fig. 2). The results are sub-optimal for other model templates, here Template B for spindown against surrounding matter beyond the ISCO satisfying $\Omega_D=\frac{1}{2}\Omega_H$, and Template C, for spindown of a PNS. 

The preferred match to Template A establishes confidence in the natural in view of spindown against matter at the ISCO, that represents 
a suspended accretion state with a luminous output in gravitational radiation \cite{van01}. A detailed analysis 
points to a sensitivity distance of about 35 Mpc by the advanced gravitational wave detectors LIGO-Virgo and KAGRA  \cite{van11a}, by extrapolating the sensitivity distance determined from model injections into the strain amplitude noise of the TAMA 300, shown in Fig. 5.

\end{multicols}
\begin{figure}[ht.]
\centerline{\includegraphics[scale=0.44]{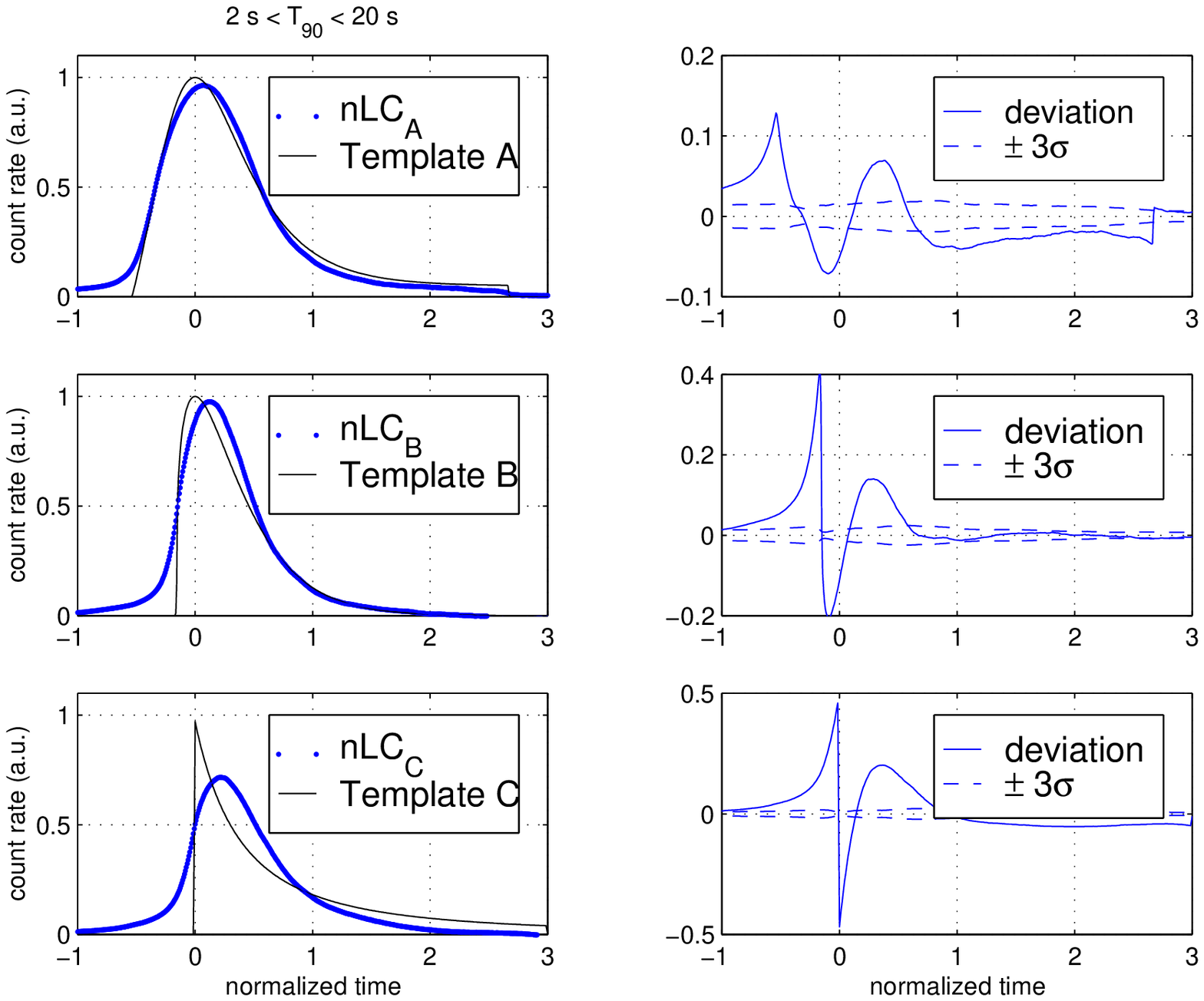}\includegraphics[scale=0.44]{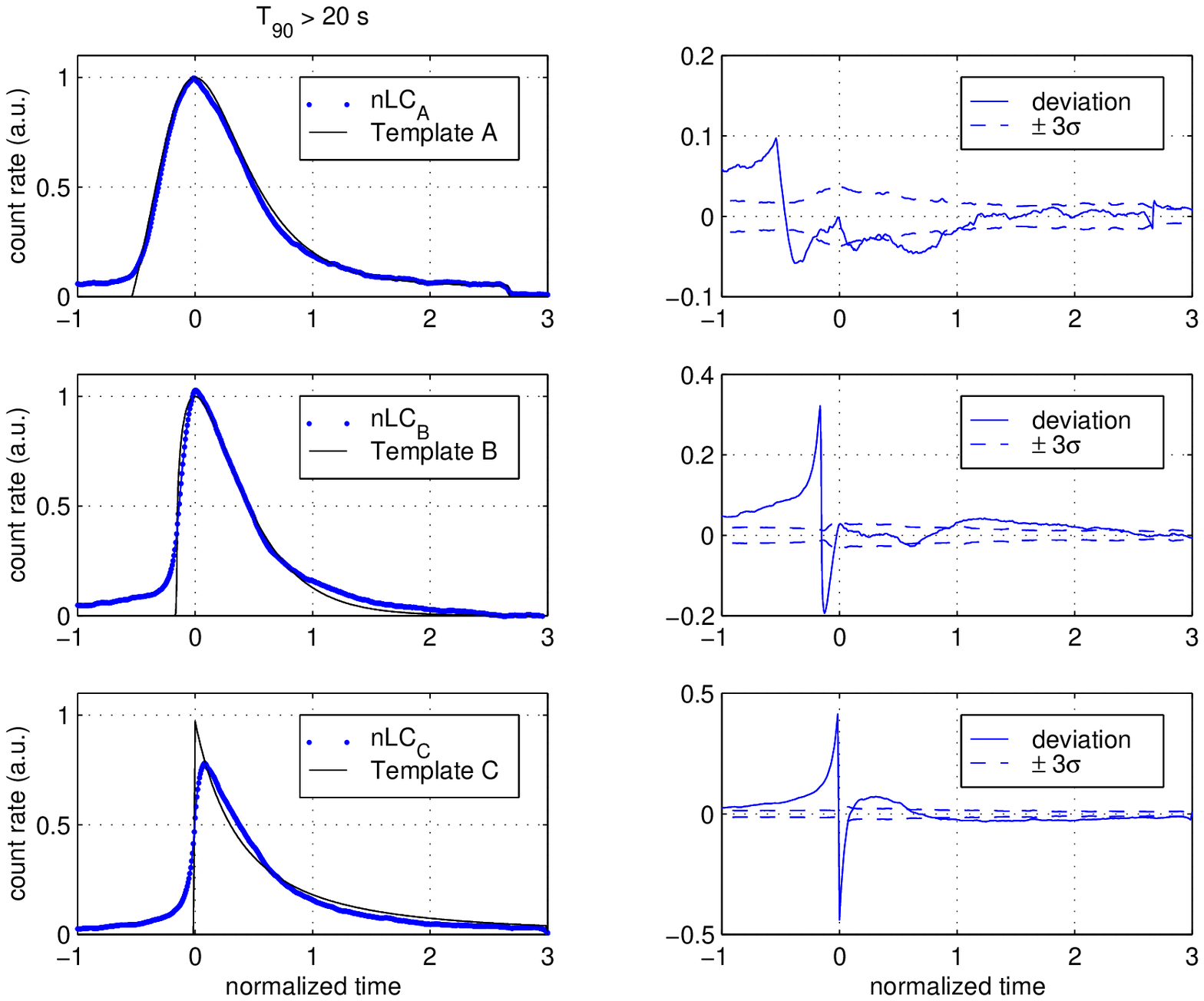}}
\caption{Shown are the nLC ($thick$ $lines$) generated by model templates AÐC ($thin$ $lines$) for the ensemble
of 531 long duration bursts with 2 s $< T_{90} <$ 20 s ($left$) and the ensemble of 960 long bursts
with $T_{90} >$20 s ($right$) and the associated deviations for Templates AÐC. Here, the standard
deviation $\sigma$ is calculated from the square root of the variance of the photon count rates in the
ensemble of individually normalized light curves as a function of normalized time. (Reprinted from \cite{van12}.)}
\label{author-fig4}
\end{figure}
\begin{multicols}{2}

\begin{myfigure}
\centerline{\resizebox{50mm}{!}{\includegraphics{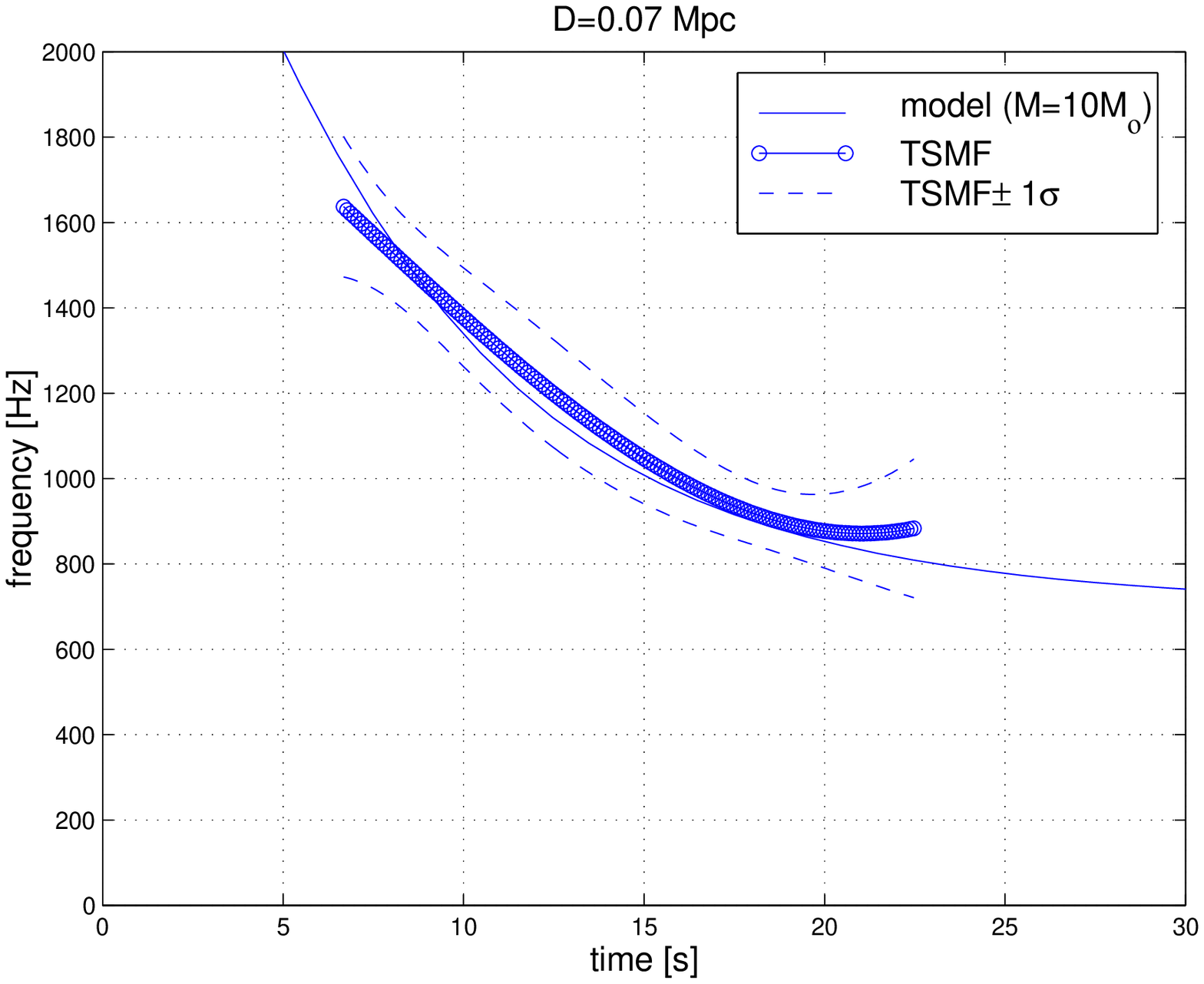}}}
\caption{The long duration negative chirp can be searched for using a dedicated time-sliced
matched filtering procedure, here illustrated following signal injection in TAMA 300 noise
data. It points to a sensitivity distance of about 35 Mpc for the upcoming advanced detectors.
(Reprinted from \cite{van11a}.)}
\label{author-fig5}
\end{myfigure}

\section{Discussion and conclusions}

Some hyper-energetic supernovae - of Type I X - appear to require an energy reservoir that exceeds the maximal rotational energy
of a PNS.  GRB-supernovae ideally has an inner engine that produces simultaneously a baryon-rich wind to power the supernova 
and a baryon-poor jet to power the GRB. These and other considerations point to a universal inner engine in the form of a rotating 
black hole, whose diversity in spin \cite{van99} can account for short/long GRBs from hyper- and suspended accretion \cite{van03}. 

The Fanaroff-Riley class I and II radio-loud AGN may represent a closely related dichotomy for supermassive black holes, where
the latter may be related to a similar suspended accretion state catalyzing black hole spin energy into powerful winds. eLISO may
test for suspended accretion by searches for an accompanying low-frequency emission in gravitational waves, notably from SgrA*. 
The Type B X-ray flaring in the galactic microquasar GRS 1915+105 may further illustrate the suspended accretion state in a 
galactic microquasar \cite{van03a}.

Fig. \ref{author-fig4} validates our model for black hole spindown most likely against matter at the ISCO, rather than further out or the spindown of (proto-)neutron stars. As a corollary, some of the neutron star-neutron star mergers are expected to long GRBs, more likely so than short GRBs. 

Our results point to specific high-frequency chirps both in GWs of interest to LIGO-Virgo and KAGRA, and
in intensity modulations in prompt GRB emissions to some of the gamma-ray satellite missions.  
For the first, we estimate a sensitivity distance to be about 35 Mpc for
the advanced generation of gravitational wave detectors, indicated in Fig. \ref{author-fig5}.

\thanks
The author thanks M. Della Valle and the organizers for the invitation to participate at this workshop and the referee
for constructive comments.

\bigskip
\bigskip
\noindent {\bf DISCUSSION}

\bigskip
\noindent {\bf J. Beal:} Is there anisotropy in the emitted gravitational radiation?

\bigskip
\noindent {\bf M. van Putten:} The anisotropy in quadrupole emissions is known to be small, about a factor of 1.58 at most.

\bigskip
\noindent {\bf B. Czerny:} (a) GW efficiency will depend strongly on the $m$ in azimuthal modes. Are low $m$ more important in the Papaloizou-Pringle instability? (b) Why don't you use the term ``hyper nova"? (c) How sensitive are your GRB templates to the detailed assumptions that go into BZ efficiency? 

\bigskip
\noindent {\bf M. van Putten:} (a) In \cite{bro06}, we considered the problem of the spectrum in gravitational wave emission using a toy model for a magnetized disk. The results indicate a dominant emission in $m=2$. In \cite{van02}, I extended the PP instability analysis to tori with finite width, and found that the low $m$ modes (starting with $m=1$) become unstable first in response to thermal or magnetic pressures. Both results are fortunate, pointing to major emission in quadrupole radiation, that falls within the limited band width of sensitivity of the LIGO-Virgo and KAGRA detectors. 

(b) In seeking EM priors on the inner engine of CC-SNe, I consider the possibility of ruling out PNS by determining the required energy reservoir relative to $E_c=3\times 10^{52}$ erg, the canonical value for the maximal energy in rotation of a PNS. The term ``hypernova" has been introduced for supernovae with isotropically equivalent kinetic energies of a few times $10^{52}$ erg, well in excess of typical SN energies of a few times $10^{51}$ erg. This definition is not sufficiently specific to serve as a prior on the inner engine. 

(c) BZ proposed an open model for a black hole luminosity entirely in open outflows, sustained by accretion of magnetized flows with no feedback onto the inner accretion disk. The black hole evolves according to the net result of losses (in energy and angular momentum) in magnetic outflow and gain by accretion. Hyper-accretion will likely cause the black hole to spin up continuously up to close to maximal rotation \cite{kum08}. In associating the duration of hyper-accretion with the observed $T_{90}$ of the bursts, a model light curve from a baryon poor jet along the black hole spin axis should be increasing, followed by a relatively sharp drop off when accretion ceases. This is at odds with the gradual and extended decay seen in the nLC obtained by Templates A-C (Fig. 4).

\end{multicols}
\end{document}